\documentclass[1p]{elsarticle}
\usepackage{graphicx}
\usepackage{caption}
\usepackage{subcaption}
\usepackage{amsmath}
\usepackage{amssymb}
\usepackage{amsthm}
\usepackage{lineno,hyperref}
\modulolinenumbers[5]
\journal{J. Magn. Magn. Mat.}

\begin{document}

\begin{frontmatter}



\title{Mixed spin-1/2 and 3/2 Ising model with multi-spin interactions on a decorated square lattice
\tnoteref{vega}}
\tnotetext[vega]{This work has been supported under grant VEGA  No. 1/0234/14 and APVV-14-0073  } 
\author[dtf]{V. \v{S}tub\v{n}a} \ead{viliamstubna@yahoo.com}
\author[dtf]{M. Ja\v{s}\v{c}ur\corref{cor1}} \ead{michal.jascur@upjs.sk}
\cortext[cor1]{Corresponding author}
\address[dtf]{Department of Theoretical Physics and Astrophysics, Institute of Physics, P.J.  \v{S}af\'{a}rik University
 in Ko\v{s}ice, Park Angelinum 9, 040 01 Ko\v{s}ice, Slovakia }
\begin{abstract}
A mixed spin-$1/2$ and spin-$3/2$ Ising model on a decorated 
square lattice with a nearest-neighbor interaction, next-nearest-neighbor bilinear interaction, three-site four-spin interaction 
and single-ion anisotropy is exactly investigated using a generalized  decoration-iteration
transformation, Callen-Suzuki identity and differential operator technique. 
The ground-state and finite-temperature phase boundaries are obtained by identifying all relevant  phases  
corresponding to minimum internal or free energy of the system.
The thermal dependencies of magnetization, correlation functions, entropy and specific heat are also calculated exactly and the most interesting cases are discussed in detail. 
\end{abstract}

\end{frontmatter}

\section{Introduction}
\label{intro}
The investigation of higher-order spin couplings  has been initiated
many decades ago by Anderson \cite{Anderson1959} and Kittel \cite{Kittel1960} who  have  studied 
the role of biquadratic exchange interactions of the form $S_i^2 S_j^2$, in connection with the superexchange 
interaction and elastic properties of magnetic materials. Later, the    higher-order spin
interactions have been experimentally found in  magnetic  compounds MnO and NiO 
\cite{Harris1963}-\cite{Rodbell1964}. 
Since then, various types of multi-spin interactions  have been intensively studied 
in  different physical systems, in order to explain  diverse physical phenomena (see for 
example \cite{Jascur2016}  and references therein). 
It has been found   that these interactions are usually much weaker than the 
standard pair Heisenberg exchange coupling, however, due to their non-conventional symmetries they may significantly modify many physical quantities in the systems under investigation.

In this work we will study a special kind of higher-order spin interactions, that are usually called  as three-site 
four-spin interactions due to their geometry. These interactions have been originally introduced  by Iwashita and Uryu 
\cite{Iwashita1974}, in order to describe magnetic properties of some clustered complex systems.  In general these interaction   take the form of $(S_iS_j)(S_jS_k)$ and  as a rule,  they principially modify physical properties of various localized magnetic system \cite{Adler1979}-\cite{Iwashita2001}. 
Among recent works in this research  field one should also notice our  study of  a decorated exactly solvable mixed 
spin-1/2 and spin-1 Ising model  with  three-site fours-spin interaction  \cite{Jascur2016}. In that work we have found  
that the such a decorated planar system may exhibit unusually interesting and rich  magnetic  behavior, among others including  also the existence of phases with non-zero ground state entropy.
Owing to many interesting phenomena found in  our previous study \cite{Jascur2016} it is of interest  to
understand the role of varying spin value in the three-site four-spin exchange interaction and to clarify which magnetic properties will be significantly changed due to varying spin of decorating  atoms.  For this purpose,  we 
will study  in this work a  bond-decorated  square lattice consisting of nodal atoms with  spin 1/2  and decorating atoms with spin 3/2.  The Hamiltonian of the systems will include, except of standard  bilinear interactions  terms, also an unconventional three-site four-spin interactions.  The role of single-ion anisotropy will be also taken into account. 
The outline of this paper is as follows. In Section 2 we briefly summarize the application of   decoration-iteration transformation to  obtain exact relations for all relevant physical quantities. In Section 3 we discuss the most interesting numerical results and finally main  conclusions are summarized in the last section. 
\section{Formulation}
The subject of our study is a mixed spin-1/2 and spin-3/2 Ising model with  bilinear,  three-site four-spin interactions and single ion anisotropy on a decorated square lattice as it is depicted in Fig 1. 
As one can see from the figure, the system consists of $N$ spin-1/2 atoms located on the sites of square lattice 
and $2N$ decorating spin-3/2 atoms that occupy all bonds of the original square lattice. 
\begin{figure}[h!]%
\centering
\includegraphics*[width = 6cm]{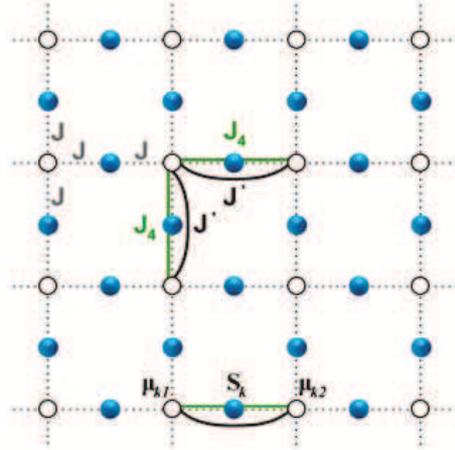}
\caption{%
 The fragment of a decorated square lattice
under investigation. The doted lines denote bilinear interactions
$J$ between nearest neighbors, the full black lines denote bilinear interactions
$J'$ between next-nearest neighbors and the green lines indicate three-site
four-spin interactions $J_{4}$. The open circles represent spin-1/2 atoms
creating the sub-lattice A and the light-blue colored ones
display decorating spin-3/2 atoms constituting the sub-lattice B. }
\label{Fig1}
\end{figure}

Thus,  as a whole the system can be treated as a two-sublattice mixed-spin system with unequal number of 
sub-lattice atoms which can be  described by the Hamiltonian
\begin{equation}
\label{eq1}
{\cal H}=\sum_{k=1}^{2N} {\cal H}_{k},
\end{equation}
where the summation runs over all bonds of the original square lattice and  $\mathcal{H}_{k}$ represents the Hamiltonian of  $k$-th bond which takes the following  explicit form
\begin{eqnarray}
\label{eq2}
\mathcal{H}_{k} & = & -J'\mu_{k1}\mu_{k2}-JS_{k}\left(\mu_{k1}+\mu_{k2}\right)\nonumber \\
 &   - &J_{4}S_{k}^{2}\mu_{k1}\mu_{k2}-DS_{k}^{2}.
\end{eqnarray}
Here the parameter $J$ denotes the bilinear exchange interaction between
nearest neighbors (n.n) $\mu_{ki}-S_{k}$, parameter $J'>0$ represents
the bilinear interaction between next-nearest neighbors (n.n.n) $\mu_{k1}-\mu_{k2}$,
the parameter $J_{4}$ stands for the three-site four-spin interaction
between $\mu_{k1}-S_{k}^{2}-\mu_{k2}$ spins and 
$D$ denotes a single-ion anisotropy. The interaction parameters $J, J_{4}$ and $D$
are allowed to take  arbitrary positive or negative values 
and the spin variables  take  obviously the following values: $\mu_{ki}=\pm 1/2$
and $S_{k}=\pm1/2,\pm 3/2$.
One should notice here that the summation in \eqref{eq1} must be performed in a way that accounts for different spin 
terms only once.

Using Eq.~ \eqref{eq1} the partition function for the present system can be written as
\begin{eqnarray}
\label{eq3}
\mathcal{Z} & = & \underset{_{\left\{ \mu_{ki}\right\} }}{\sum}
\underset{_{\left\{ S_{k}\right\} }}{\sum}\exp\left(-\beta\overset{2N}{\sum_{k=1}}\mathcal{H}_{k}\right)\nonumber \\
 & = & \underset{_{\left\{ \mu_{ki}\right\} }}{\sum}\prod_{k}^{2N}\sum_{S_{k}=\pm\frac{1}{2},\pm\frac{3}{2}}\exp\left(-\beta\mathcal{H}_{k}\right),
\end{eqnarray}
where $\beta=1/k_{B}T$, $k_{B}$ is the Boltzmann constant,
$T$ is absolute temperature and lastly $\sum_{\left\{ \mu_{ki}\right\} }$
and $\sum_{\left\{ S_{k}\right\} }$ mean summation over
degrees of freedom of $\mu_{ki}$ and $S_{k}$ spins, respectively.

Now introducing the following generalized decoration-iteration transformation \cite{Syozi1952}-\cite{Jascur2001}.
\begin{equation}
\label{eq4}
\sum_{ S_{k}=\pm\frac{1}{2},\pm\frac{3}{2} }\exp (-\beta\mathcal{H}_{k} )
= Ae^{\beta R\mu_{k1}\mu_{k2}},
\end{equation}
one may recast the partition function of the system in the form
\begin{equation}
\mathcal{Z}=A^{2N}\mathcal{Z}_{0}(\beta R)
\label{eq5}
\end{equation}
where $\mathcal{Z}_{0}$ represents the partition function of the original (undecorated)  square lattice with
$N$ spin-1/2 atoms   that interact via nearest-neighbor  effective exchange interaction $R$. Here one should recall that 
exact analytical expresion of $\mathcal{Z}_{0}$ is well known from the Onsager seminal work \cite{Onsager1944}. 
The expression $A^{2N}$ represents the contribution of the decorating spins to the total partition function.
  
Both unknown parameters $A$, $R$ entering Eq.~\eqref{eq5} can be straightforwardly evaluated performing the 
summation on the l.h.s. in Eq.~\eqref{eq4} and substituting   $\mu_{k1}=\pm1/2$ and $\mu_{k2}=\pm 1/2$ 
into resulting expression. In this way one easily gets the following relations
\begin{eqnarray}
A & = & \left(V_{1}V_{2}\right)^{\frac{1}{2}},
\label{eq6}\\
\beta R & = & 2\ln\frac{V_{1}}{V_{2}},
\label{eq7}
\end{eqnarray}
where 
\begin{eqnarray}
V_{1} & = & 2e^{\frac{1}{4}\beta J'}e^{\frac{9}{16}\beta J_{4}}e^{\frac{9}{4}\beta D}K_{1},
\label{eq8}\\
V_{2} & = & 2e^{-\frac{1}{4}\beta J'}e^{-\frac{9}{16}\beta J_{4}}e^{\frac{9}{4}\beta D}K_{2},
\label{eq9}
\end{eqnarray}
with
\begin{eqnarray}
K_{1} & = & \cosh\left(\frac{3}{2}\beta J\right)+
\mbox{e}^{-\frac{1}{2}\beta J_{4}}e^{-2\beta D}\cosh\left(\frac{1}{2}\beta J\right)
\label{eq10}\\
K_{2} & =& 1+e^{\frac{1}{2}\beta J_{4}}\mbox{e}^{-2\beta D}.
\label{eq11}
\end{eqnarray}
Having obtained exact mapping relations \eqref{eq6}-\eqref{eq11}, and exact expression for the partition function 
of the system \eqref{eq5}, we are now able  to gain exact equations for  phase boundaries and exact analytical relations for  all physical quantities of interest.

At first, after substituting the value of inverse critical temperature of the square lattice,    
$\beta_{c}R=2\ln\left(1+\sqrt{2}\right)$,  into l.h.s of \eqref{eq7}, we obtain formula for finite-temperature phase diagrams of the decorated system in the form
\begin{equation}
1 + \sqrt{2} = \frac{V_{1c}}{V_{2c}}, 
\label{eq12}
\end{equation}
where $V_{1c} = V_1(\beta_c)$,  $V_{2c} = V_2(\beta_c)$ and  $\beta_c = 1/k_B T_c$.

Next, using the relation $F = -k_BT \ln \mathcal{Z} $ one simply obtains from Eq-~\eqref{eq5}
the following relation for the Helmholtz free energy of the entire system
\begin{eqnarray}
\nonumber
  F( \beta, J,  J_4,  J',  D)  = &-& 2N \beta^{-1} \ln A ( \beta, J,  J_4,  J',  D) 
  \\
       &+& F_0(\beta, R),
  \label{eq13}
\end{eqnarray}
where parameters  $ A $,  $R $ are given by Eq.~\eqref{eq6}-\eqref{eq7} and $ F_0(\beta, R) $ represents the Helmholtz free energy  of the original undecirated  Ising  square lattice \cite{Onsager1944}
\begin{equation}
F_{0}=-\frac{N}{\beta }\left[\ln\left(2\cosh\frac{\beta R}{2}\right)+\frac{1}{2\pi}\intop_{0}^{\pi}\zeta\left(\phi\right)d\phi\right],
\label{eq14}
\end{equation}
with
\begin{equation}
\zeta\left(\phi\right)=\ln\left[\frac{1}{2}\left(1+\sqrt{1-\kappa^{2}\sin^{2}\phi}\right)\right],
\end{equation}
and
\begin{equation}
\label{eq15}
 \kappa = \frac{2\sinh \bigl (\frac{\beta R}{2} \bigr) }{  \cosh^2 \bigl( \frac{\beta R}{2} \bigr )}.
\end{equation}
Now, the  entropy $S$  and specific heat  $C$  can calculated from Eq.~\eqref{eq13} using the following  
thermodynamic relations
\begin{equation}
S=-\left(\frac{\partial F}{\partial T}\right)_V,  \qquad 
C_V= -T\left(\frac{\partial^2 F}{\partial T^2}\right)_V,
\label{eq17}
\end{equation}
and consequently the internal energy can be also easily obtained using equation
\begin{equation}
U = F + TS.
\label{eq18}
\end{equation}
In addition to the above mentioned thermodynamic quantities, we will also investigate
the spin-ordering  in all possible phases of the system. For this purpose it is necessary  to analyze   the total and sub-lattice magnetization along with various spin-correlation functions.

The total magnetization per one site of the decorated lattice is given by 
\begin{equation}
                   m=\left(m_{A}+2m_{B}\right)/3, 
\label{eq19}
\end{equation}
where  the sub-lattice magnetization  $m_{A}$ and   $m_{B}$ are respectively given by 
\begin{equation}
m_A=\left\langle \mu_{ki}\right\rangle  = \frac{1}{\mathcal{Z}}\underset{_{\left\{ \mu_{ki}\right\} }}{\sum}
\underset{_{\left\{ S_{k}\right\} }}{\sum}\mu_{ki}\exp\left(-\beta\overset{2N}{\sum_{k=1}}\mathcal{H}_{k}\right)
\label{eq20}
\end{equation}
\begin{equation}
m_B=\left\langle S_{k}\right\rangle  = \frac{1}{\mathcal{Z}}\underset{_{\left\{ \mu_{ki}\right\} }}{\sum}
\underset{_{\left\{ S_{k}\right\} }}{\sum}S_{k}\exp\left(-\beta\overset{2N}{\sum_{k=1}}\mathcal{H}_{k}\right)
\label{eq21}
\end{equation}
The calculation of $m_A$ is a particularly simple task, since using   Eqs.~\eqref{eq4} and \eqref{eq5} one obtains
from \eqref{eq20} the following relation \cite{Jascur2001},  \cite{Barry1988}, \cite{Barry1991}
\begin{equation}
\left\langle f(\mu_{k1}, \mu_{k2}, \dots, \mu_{ki})\right\rangle  =
\left\langle f(\mu_{k1}, \mu_{k2}, \dots, \mu_{ki})\right\rangle_0. 
\label{eq22}
\end{equation}
Here $f$ represents an arbitrary function depending exclusively on the spin variables of  A sublattice.  
Thus, setting $f(\mu_{k1}, \mu_{k2}, \dots, \mu_{ki}) = \mu_{ki} $ one obtains
$ \left\langle \mu_{ki}\right\rangle = \left\langle \mu_{ki}\right\rangle_0 = m_0 $, where  $m_0$ represents the 
magnetization per one lattice site of the original square lattice which has been exactly calculated by Yang \cite{Yang1952} and in our case it takes the form 
\begin{equation}
m _{0}=\frac{1}{2}\left(1-\frac{16 e^{-2\beta R}}{\left(1-e^{-\beta R}\right)^{4}}\right)^{\frac{1}{8}}.
\label{eq23}
\end{equation}
On the other hand,  for the calculation of $m_B$ i.e., the mean value of $\left\langle S_{k}\right\rangle $, one can use
the exact Callen-Suzuki identity \cite{Callen1963}, \cite{Suzuki1965}, \cite{Jascur2015} as a starting point.
\begin{equation}
\left\langle S_{k}\right\rangle =\left\langle \frac{\sum_{S_k}S_{k}\mbox{e}^{-\beta\mathcal{H}_{k}}}{\sum_{S_k} \mbox{e}^{-\beta\mathcal{H}_{k}}}\right\rangle ,
\label{eq24}
\end{equation}
where $\mathcal{H}_{k}$ is defined in Eq. \eqref{eq2}.
After performing the summation over $S_k$ in previous equation one obtains 
\begin{equation}
\left\langle S_{k}\right\rangle =\frac{1}{2}\left\langle \frac{3\sinh\left(\frac{3 \beta}{2} h\right)
   + \mbox{e}^{-2\beta(h_4 +D)}\sinh\left(\frac{\beta}{2}h\right)}
  {\cosh\left(\frac{3 \beta}{2} h\right) + \mbox{e}^{-2\beta(h_4 +D)}\cosh\left(\frac{\beta}{2}h\right)}\right\rangle ,
\label{eq25}
\end{equation}
where we have denoted effective fields acting on the $k$-th lattice site as
\begin{equation}
h = J(\mu_{k1} + \mu_{k2}),  \quad h_4 = J_4 \mu_{k1}  \mu_{k2}.
\label{eq26}
\end{equation}
In order to calculate the ensemble average in last equation it is very comfortable to utilize
the  differential operator method which is based on the following relations 
\begin{equation}
f\left(x +\lambda_{x}, y + \lambda_{y}\right)
=\mbox{e}^{\left(\lambda_{x}\nabla_{x}+\lambda_{y}\nabla_{y}\right)}f\left(x,y\right)
\label{eq27}
\end{equation}
\begin{equation}
\mbox{e}^{a \mu}=\cosh\left(\frac{a}{2}\right)+2\mu\sinh\left(\frac{a}{2}\right), \quad \mu = \pm 
\frac{1}{2}
\label{eq28}
\end{equation}
where $\nabla_{x}=\partial/\partial x$, $\nabla_{y}=\partial/\partial y$ are standard differential operators and 
$a$ stands for an arbitrary parameter.

Now, with the help of  \eqref{eq27}  and   \eqref{eq28}  one obtains for the sublattice magnetization $m_B$ the  
following simple expression

\begin{equation}
\left\langle S_{k}\right\rangle =\left\langle \mu_{k1}\right\rangle A_{1},
\label{eq29}
\end{equation}
where 
\begin{equation}
A_{1} = \frac{3\sinh\left(\frac{3}{2}\beta J\right)+\mbox{e}^{-\frac{1}{2}\beta J_4}\mbox{e}^{-2\beta D}
\sinh\left(\frac{1}{2}\beta J\right)}
{\cosh\left(\frac{3}{2}\beta J\right)+
\mbox{e}^{-\frac{1}{2}\beta J_{4}}\mbox{e}^{-2\beta D}\cosh\left(\frac{1}{2}\beta J\right)}.
\label{eq30}
\end{equation}
Subsequently, the total magnetization per one lattice site \eqref{eq19}
for the present system can be explicitly written in the form
\begin{equation}
m=\frac{1}{3}\left\langle \mu_{k1}\right\rangle \left(1+2A_{1}\right).
\label{eq31}
\end{equation}
Hawing obtained the total reduced magnetization, one can now calculate the compensation temperature\textbf{ 
$T_{k}$} from the condition $m=0\,\wedge\, m_{A}\neq0\,\wedge\, m_{B}\neq0$ \cite{Neel1948}.
By means of this definition, we derive the condition for \textbf{$T_{k}$} in the following form
\begin{equation}
7w^{\frac{3}{2}\alpha}-5w^{-\frac{3}{2}\alpha}+3w^{\gamma_{-}}-w^{\gamma_{+}}=0
\label{eq32}
\end{equation}
where we have defined the following terms
\begin{eqnarray*}
w & = & e^{\beta_{k}J_{4}}\\
\gamma_{+} & = & -\frac{1}{2}\left(4d+1+\alpha\right)\\
\gamma_{-} &= & -\frac{1}{2}\left(4d+1-\alpha\right).
\end{eqnarray*}
Of course, the equation \eqref{eq32} has to be used in
line with inequality $T_{k}<T_{c}$.
In a similar way, one also obtains equations for the quadrupolar moment and various spin-correlation functions:
\begin{eqnarray}
q_B = \left\langle S_{k}^{2}\right\rangle  & = & \frac{1}{8}\left(A_{2}+A_{3}\right)+\frac{1}{2}\left\langle \mu_{k1}\mu_{k2}\right\rangle \left(A_{2}-A_{3}\right),
\label{eq33}
\nonumber \\ \\ 
\left\langle S_{k}\mu_{k1}\right\rangle  & = & \left(\frac{1}{8}+\frac{1}{2}\left\langle \mu_{k1}\mu_{k2}\right\rangle \right)A_{1},
\label{eq34}
\\
\left\langle S_{k}^{2}\mu_{k}\right\rangle  & = & \frac{1}{4}\left\langle \mu_{k}\right\rangle A_{2},
\label{eq35} 
 \\ \nonumber
\left\langle S_{k}^{2}\mu_{k1}\mu_{k2}\right\rangle  & = & \frac{1}{32}\left(A_{2}-A_{3}\right)+\frac{1}{8}\left\langle \mu_{k1}\mu_{k2}\right\rangle \left(A_{2}+A_{3}\right), 
\label{eq36}
\\
\end{eqnarray}
where the coefficients  $A_{2}$ and $A_{3}$ are defined as
\begin{eqnarray}
A_{2} & = & \frac{9\cosh\left(\frac{3}{2}\beta J\right)+e^{-\frac{1}{2}\beta J_{4}}e^{-2\beta D}\cosh\left(\frac{1}{2}\beta J\right)}{\cosh\left(\frac{3}{2}\beta J\right)+e^{-\frac{1}{2}\beta J_{4}}e^{-2\beta D}\cosh\left(\frac{1}{2}\beta J\right)},
\label{eq37}\\
A_{3} & = & \frac{9+e^{\frac{1}{2}\beta J_{4}}e^{-2\beta D}}{1+e^{\frac{1}{2}\beta J_{4}}e^{-2\beta D}}.
\label{eq38}
\end{eqnarray}
We recall here that  average values  entering r.h.s. of \eqref{eq29}, \eqref{eq33}-\eqref{eq36} can be simply 
evaluted,  since on the basis of Eq. \eqref{eq22} one obtains $\left\langle \mu_{k1}\right\rangle = \left\langle \mu_{k1}\right\rangle_0$ and $\left\langle \mu_{k1}\mu_{k2}\right\rangle = \left\langle \mu_{k1}\mu_{k2}\right\rangle_0$.

\section{Numerical results}
In this part we will present the most interesting  results obtained numerically from equations presented in the previous section.  For the sake of simplicity,  we introduce the following dimensionless parameters
 $\alpha= J/J_4$,  $d = D/J_4$,  $\lambda = J^{\prime}/J_4$.
\subsection{Ground-state  phase diagrams}
\begin{figure}[h!]%
\centering
\includegraphics*[width=8cm]{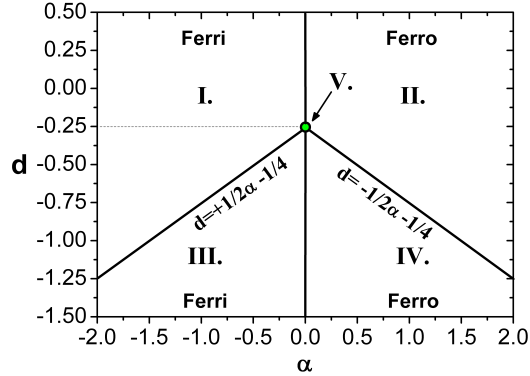}
\caption{%
 The ground-state phase diagram in $\alpha-d$
space for arbitrary $\lambda$. In different sectors there are phases with different spin configurations indicated in the figure. }
\label{Fig2}
\end{figure}
The ground-state phase diagram of the present system has been obtained by investigating the internal energy
of all relevant spin configurations at $T=0$. Our findings are summarized in Fig.~\ref{Fig2} where we have  
depicted the phase diagram  in $\alpha-d$ plane   which is valid for arbitrary $\lambda$. As one can see,  the 
whole parameter space is divided into four  regions in which  different ordered magnetic phases can be found.
 Namely:\\
I.   the ferrimagnetic phase with  $m_A=1/2, m_B =  -3/2$  and $q_B = 9/4$,
     for $\alpha < 0$ and $d > \alpha/2 - 1/4$. 
\\
II.  the ferromagnetic phase with  $m_A=1/2, m_B =  3/2$  and $q_B = 9/4$,
     for $\alpha > 0$ and $d >  -\alpha/2 - 1/4$.  
\\ 
III.  the ferrimagnetic phase with  $m_A=1/2, m_B =  -1/2/2$  and $q_B = 1/4$,
      for $\alpha < 0$ and $d <  \alpha/2 - 1/4$. 
\\
IV.   the ferromagnetic phase with  $m_A=1/2, m_B =  1/2$  and $q_B = 1/4$,
     for $\alpha > 0$ and $d < -\alpha/2 - 1/4$.  
\\
The boundaries  separating  these regions represent the lines of first-order phase transitions along which 
relevant couples of phases co-exist.  Therefore, for $d = \alpha/2 - 1/4$ and $\alpha<0$,  we have found the 
phase with $m_A=1/2, m_B =  -1$  and $q_B = 5/4$,  while for  $d = -\alpha/2 - 1/4$ and $\alpha>0$,  one 
finds $m_A=1/2, m_B =  1$  and $q_B = 5/4$.  Similarly,  for the case of pure three-site four spin interaction,
(i.e. $\alpha=0$),  we have again  the co-existence of relevant phases, but now the resulting  phase will be only 
partially ordered, since each decorating atom occupies  equally likely $\pm 3/2$ or $\pm 1/2$ spin states, respectively. 
Consequently, for $\alpha=0$ and $d>-1/4$ one gets $m_A=1/2, m_B =  0$  and $q_B = 9/4$, while for 
$\alpha=0$ and $d<1/4$  one finds $m_A=1/2, m_B =  0$  and $q_B = 1/4$. Finally, the  point with 
co-ordinates ($\alpha, d$) = (0, -1/4) represents a special point in which coexist all four ordered phases, so that
the minimum of the internal energy now corresponds to the partially ordered phase with $m_A=1/2, m_B =  0$  
and $q_B = 5/4$, since now all the states  $S_k =\pm 1/2,  \pm 3/2$ on B sublattice are occupied equally likely. 
Here one should notice that the disorder  appearing along  the phase boundary $\alpha=0$ will 
naturally generate non-zero values of the entropy at the ground state. This interesting phenomenon will be 
discussed in detail in Subsection 3.2. 
\begin{figure}[h!]%
\centering
\includegraphics*[width=8cm]{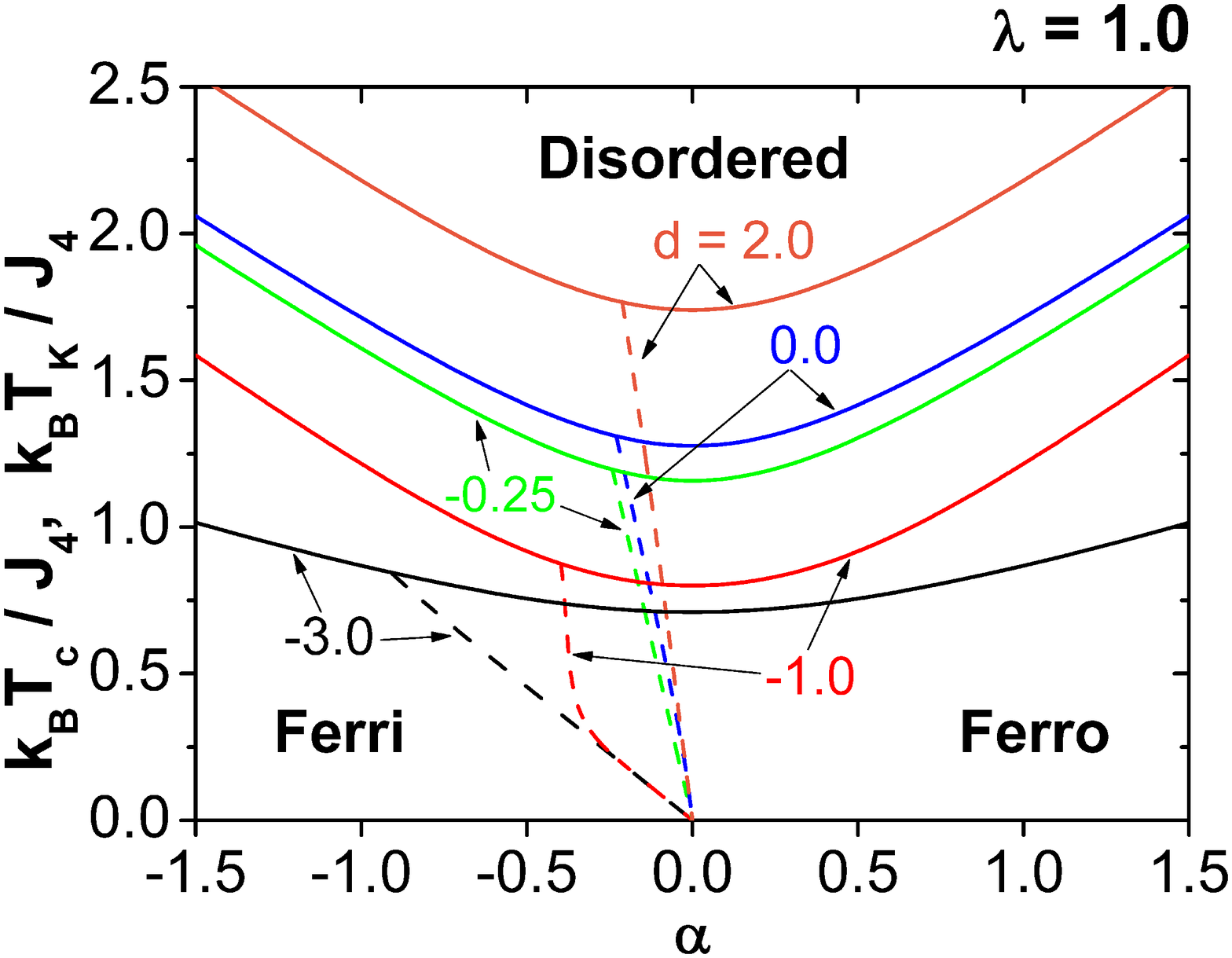}
\caption{%
The  critical (solid curves) and compensation temperatures (dashed curves)  in $\alpha-T$ space for 
$\lambda=1.0$ and different values of $d$.}
\label{Fig3}
\end{figure} 
\subsection{Finite-temperature phase diagrams and thermodynamic properties}
\begin{figure}[h!]%
\centering
\includegraphics*[width=8cm]{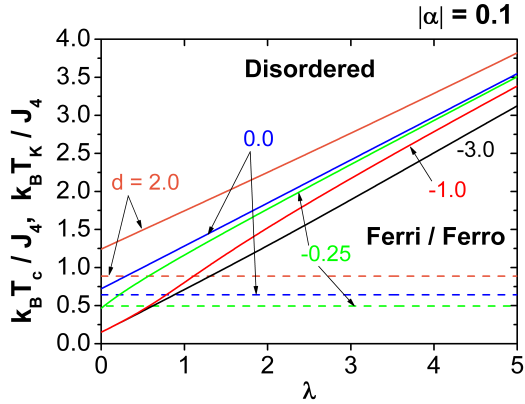}
\caption{%
The  critical (solid curves) and compensation temperatures (dashed curves)  in $\lambda-T$ space for 
$|\alpha| = 0.1$ and different values of $d$. The critical boundaries are identical for ferromagnetic ($\alpha =0.1$) and
ferrimagnetic ($\alpha = -0.1$) cases, while the compensation temperatures are possible only for ferrimagnetic phase 
($\alpha = -0.1$).}
\label{Fig4}
\end{figure}

\begin{figure}[h!]%
\centering
\includegraphics*[width=8cm]{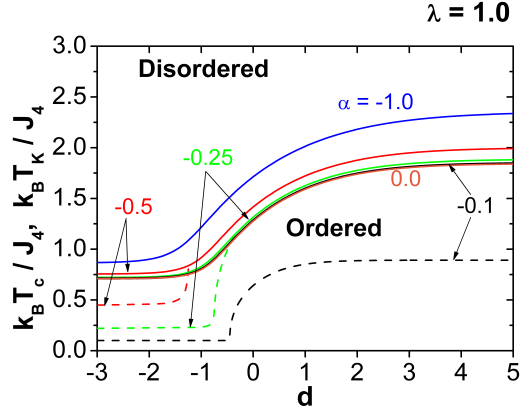}
\caption{%
The  critical (solid curves) and compensation temperatures (dashed curves)  in $d-T$ space for 
$\lambda=1.0$ and different values of $\alpha$.}
\label{Fig5}
\end{figure}

In order to investigate  thermal properties of the system,   we have at first 
calculated  critical and compensation temperatures  using Eq. \eqref{eq12} and Eq. \eqref{eq32}.   In Figs. \ref{Fig3}-\ref{Fig5}  we 
have depicted  representative results by selecting various combinations of model parameters. At 
first,  in Fig. 3  we have shown the results in the $\alpha - T_c$ space for $\lambda = 1.0$  and some 
characteristic values of the single-ion anisotropy parameter $d$. In the figure, the solid and doted curves correspond 
to  critical and compensation temperatures, respectively.  In agreement with the ground-state analysis,  one  finds 
the standard  ferri- or ferromagnetic phases to be  stable at low temperatures for $\alpha<0$ or $\alpha>0$, 
respectively.  On the other hand, the standard paramagnetic phase exists above each phase boundary. 
It is therefore clear  that the normalized spontaneous magnetization of these phases will take its 
saturation value at $T=0$ and then  will gradually decrease with increasing temperature,   until it  continuously 
vanishes at the  corresponding critical temperature.  Moreover one can see that for fixed values of $\lambda$ 
and $d$ one can observe the compensation effect by choosing  appropriate negative values of the parameter 
$\alpha$.    
We have also investigated  other cases and we have found  that independently of the values of $d$ and $\lambda$ 
all phase boundaries exhibit a symmetric U-shape form with minimum values at $\alpha= 0$. Here one should recall 
that the phase boundary for $\alpha = 0$ represents  the critical temperatures of partially ordered phase. 
Intuitively one can simply  understand the minimum value of $T_c$, since in the partially ordered phase (i.e. for 
$\alpha = 0$) only the sublattice A  exhibits a non-zero magnetization and therefore in this case  it is  easier   to 
destroy the long-range order than that one appearing in  the fully ordered  ferrimagnetic or ferromagnetic  phases. 
Next, in order to demonstrate the influence of the n.n.n. bilinear interaction on physical behavior of our system, we 
have studied  critical boundaries and  compensation temperatures in  $\lambda - T$ space. 
Our findings are summarized in Fig. \ref{Fig4} for  $|\alpha| = 0.1$ and characteristic values of the parameter $d$.
As expected, all phase boundaries exhibit almost a perfect linear dependence   with the increasing strength of the 
parameter $\lambda$ and the compensation temperatures are again clearly visible for $\alpha = -0.1$ and 
appropriate combinations of the parameters $\alpha $ and $d$.

Finally, we have studied the critical and compensation temperatures in $d - T_c$ space. As one can see from Figs. 5 
 the phase boundaries have very similar shapes as a phase diagram of  the standard spin-3/2 Blume-Capel 
model \cite{Kaneyoshi1992}. This typical behavior of the system is mainly driven by the variation of the crystal-field 
parameter $d$ and it is clear that on the sublattice B positive values of  $d$ promote  the spin states $\pm 3/2$,  
while the negative values prefer the occupation of $\pm1/2$ spin states.  One should emphasize here that such  
suppressing or favoring of the relevant spin states on the B sublattice has a principal influence on the three-site 
four-spin interaction term which which takes the form of  $- J_{4}S_{k}^{2}\mu_{k1}\mu_{k2}$. 
As a matter of fact,   one easily identifies that at the ground state  this three-site term reduces to 
$- 9J_{4}\mu_{k1}\mu_{k2}/4$ or $- J_{4}\mu_{k1}\mu_{k2}/4$ for $d \to \infty$ or $d \to -\infty$, respectively.
Thus it is clear that for strong values of the crystal field the three-site four-spin  interaction  acts as an effective
n.n.n pair interaction which substantially reinforces the effect of parameter $\lambda$ and consequently keeps
high values of the sublattice magnetization $m_A$ even at higher temperature region. Due to this interesting effect, 
the compensation temperatures may exist in very large regions of $d$ and they can exhibit very interesting  
behavior. In fact, we have found that for a non-zero   $\lambda$ and some appropriate values of $\alpha$ the 
compensation temperature always  takes its saturation value for   $d \to -\infty$ and it very slowly increases with 
increasing the crystal-field parameter. On the other hand, the change of relevant compensation curves becomes 
more  dramatic in the region of $ d > -2.0$, where for $\alpha < -0.211$ each compensation temperature terminates 
at the relevant critical boundary, while for $\alpha > -0.211$ the existence of compensation temperatures extends up to infinite values of the crystal-field parameter.
\begin{figure}[h!]%
\centering
\includegraphics*[width=8cm]{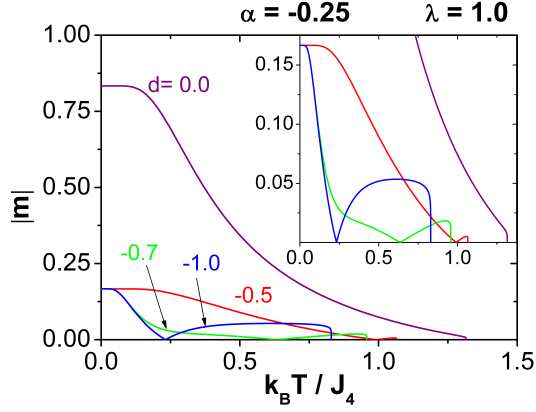}
\caption{%
Thermal variations of the absolute value of total magnetization for $\alpha = -0.25$, $\lambda = 1.0$ and several characteristic values of the parameter $d$.}
\label{Fig6}
\end{figure}
Numerical analysis of the critical and compensation temperatures  has clearly revealed several interesting 
physical phenomena that appear in our system due to the presence of unconventional three-site four-spin
interaction. To put further insight on thermodynamic properties of the system,   let us discuss the most 
interesting thermal variation of the magnetization, entropy, specific heat and Helmholtz free energy.
As far as concerns the magnetization, we restrict ourselves  to the ferrimagnetic case, in order to illustrate 
existence of compensation points at finite temperatures.  For this purpose,  in Fig. \ref{Fig6},  we have depicted 
several temperature dependencies of the absolute value of total magnetization per one atom by selecting suitable 
combinations of all 
relevant parameters. The presented curves  are in perfect agreement with our results presented in Figs. 2-5.  Here 
one should mention that the detailed theoretical investigation of  magnetic  systems exhibiting compensation 
temperatures  is also of interest in connection with development of new recording media.  
\begin{figure}[h!]%
\centering
\includegraphics*[width=8cm]{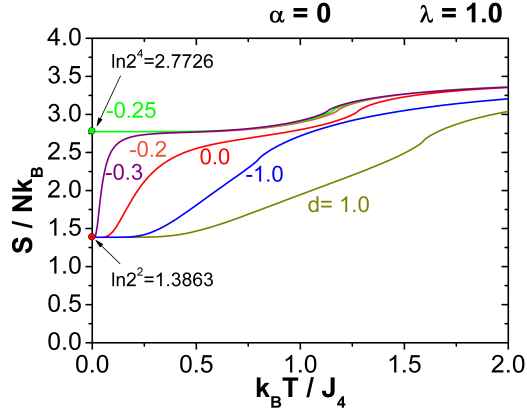}
\caption{%
Thermal variations of the reduced magnetic entropy for  $\alpha = 0$, $\lambda = 1.0$ and several characteristic values of the parameter $d$.}
\label{Fig7}
\end{figure}

Now  let us turn our attention to the analysis of the thermal variations of entropy and specific heat. 
Here we will  preferably  discuss the  partially ordered phase i.e. $\alpha = 0$ and also
 the phases  that are stable along the phase boundaries given by  equations $d = \pm 0.5 \alpha - 0.25 $. 
As we have already mentioned above, in the case of $\alpha = 0$ the sublattice B remains 
strongly disordered down to zero temperature, while the sublatice A  exhibits the standard long-range order
at $T=0$, thus the non-zero  entropy appears at the ground state. The situation is shown in Fig. \ref{Fig7}, where we have 
presented the thermal variations  of the entropy for $\alpha = 0$ and $\lambda = 1.0$ and several typical negative and 
positive values of $d$. As one can see from the figure,   for $d=-0.25$ we have obtained $S/Nk_B = \ln (2^4) = 2.7726 $ while 
for all other values of $d \neq -0.25$ one finds $S/Nk_B = \ln (2^2) = 1.3863$.  Here one should notice that for $\alpha =0$ 
remain the ground-state values of the entropy  unchanged even for arbitrary non-negative $\lambda$. 
\begin{figure}[h!]%
\centering
\includegraphics*[width=8cm]{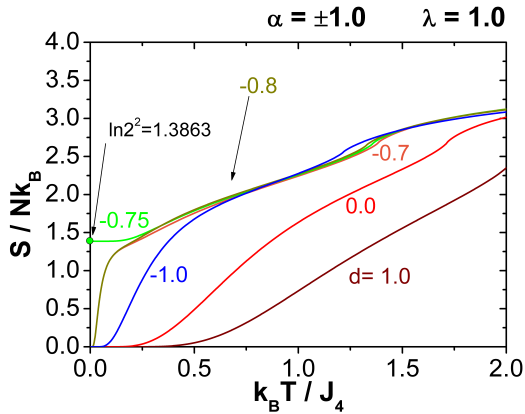}
\caption{%
Thermal variations of the reduced magnetic entropy for  $\alpha = \pm 1$, $\lambda = 1.0$ and several characteristic values of the parameter $d$.}
\label{Fig8}
\end{figure}

Next,  in order to demonstrate the role of n.n, pair interaction,  we have in Fig. \ref{Fig8} depicted temperature dependencies 
of the entropy for $\alpha = \pm 1$, $\lambda =1.0$ and some generic values of $d$. In this case only one non zero 
value of the entropy appears for $d = -0.75$, which corresponds to the ground-state phase point located exactly on 
the line given by the equation $d = \pm 0.5 \alpha - 0.25$ and again, this situation does not change for non-negative
values of $\lambda$.    
Moreover,  our results also indicate that  the entropy does not depend on the sign of parameter $\alpha$, thus for  
arbitrary  fixed values of $\lambda$ and $d$,  the system takes  exactly the same values of   entropy for the 
ferromagnetic  as well as  for the ferrimagnetic equilibrium thermodynamic states.    Of course, the described 
behavior of entropy is in a full agreement with our previous discussion.  
\begin{figure}[h!]%
\centering
\includegraphics*[width=8cm]{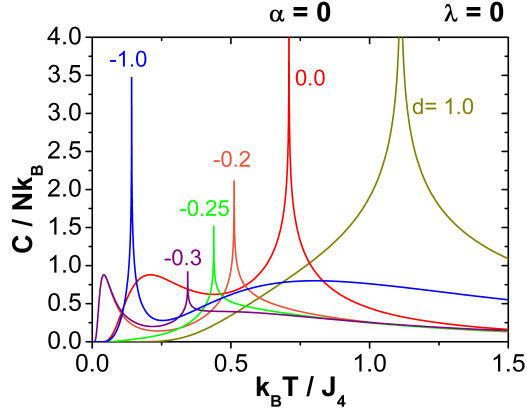}
\caption{%
Temperature dependencies of the reduced magnetic specific heat for $\alpha = 0$, $\lambda = 0$  and several characteristic values of the parameter $d$.}
\label{Fig9}
\end{figure}
\begin{figure}[h!]%
\centering
\includegraphics*[width=8cm]{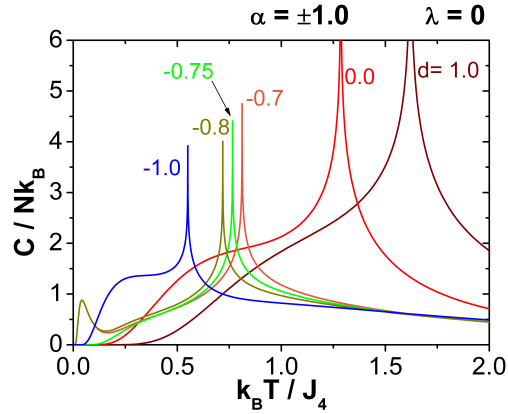}
\caption{%
Temperature dependencies of the reduced magnetic specific heat for $\alpha = \pm 1$, $\lambda = 0$  and several characteristic values of the parameter $d$.}
\label{Fig10}
\end{figure}
To complete our investigation of thermal properties of our system we have calculated the magnetic part of the specific 
heat. Our main finding are illustrated in Fig.\ref{Fig9} and  Fig. \ref{Fig10} for several representative combinations of parameters. As one 
can see, all curves  go to zero value for $T \to 0$ in agreement with the Third Law of Thermodynamics and also each 
curve exhibits Onsager type singularity at the corresponding critical temperature. We can also observe that several 
thermal dependencies of the specific heat exhibit a very clear local maximum at low-temperature region. This phenomenon is 
observable whenever the relevant set of parameters is selected from the close neighborhood  of ground-state phase boundaries. In such a case there appears a strong mixing of different spin states  on the B sublattice, since the 
relevant stets are very easily thermally excited.

\section{Conclusion}
In this paper we have investigated phase diagrams and thermal properties of the complex mixed spin-1/2 and 
spin-3/2 Ising model on decorated square lattice.   We have mainly concentrated on understanding of the influence of   
three-site four-spin interaction on magnetic properties of the system. Applying  the generalized decoration-iteration 
transformation we have obtained exact results for phase diagrams and all relevant thermodynamic quantities of the 
model. We   have also clearly demonstrated that due to multi-spin interactions the model exhibits  several unexpected features, 
for example, the existence of a partially ordered phase or non-zero ground state entropy. Comparing the present 
results with those obtained in our previous work for the decorating spin value of $S_B =1$,  one can then formulate  
the following general physical statements:\\
1. On the contrary to four-site four-spin coupling,  the three-site four spin interaction is able to generate a partially 
ordered phase even in the magnetic systems without bilinear interactions. This partially ordered phase can be stable 
in a wide temperature region and it exhibits a second-order phase transition at some critical temperature, which can depend on other physical parameters of the model, such as the crystal field.\\
2. Due to its special symmetry with respect to spatial reversal of spins,  the three-site four-spin interaction always 
suppresses  the long-range ordering in arbitrary magnetic phase.\\
3. In the mixed-spin  magnetic systems with three-site four-spin interactions the paramagnetic phase can by stable at 
$T=0$ for negative values of the crystal field, whenever atoms of the one sublattice are integer (i.e. $S_B = 1,2,... 
$). This behavior is impossible to observe in the systems with a half-integer values of $S_B$. \\

In general,  the theoretical investigation of the systems with many-body interactions is extraordinarily  
complicated task, however, the localized-spin models represent an excellent  basis for deep understanding of various 
many-body interactions going beyond the standard pair-wise  picture.  For that reason, we hope that the present 
study may initiate a wider interest in investigation of magnetic systems with multi-spin interactions.

\bibliographystyle{model1-num-names}

\end{document}